\documentstyle[aps,pra,twocolumn,epsfig]{revtex}

\draft

\begin{document}
\title{Quantum homodyne tomography with {\em a priori} constraints}
\author{Konrad Banaszek}
\address{Instytut Fizyki Teoretycznej, Uniwersytet Warszawski,
Ho\.{z}a 69, PL-00-681 Warszawa, Poland}
\date{\today}

\maketitle

\begin{abstract}
I present a novel algorithm for reconstructing the Wigner function from
homodyne statistics. The proposed method, based on maximum-likelihood
estimation, is capable of compensating for detection losses in a
numerically stable way.
\end{abstract}


An intriguing aspect of quantum mechanics is the intricate
description of the state of a physical system. Therefore, a
great deal of interest has been attracted by the recent experimental
demonstration that the quantum state of a simple system, namely
a single light mode, can be completely characterized in a feasible
scheme \cite{SmitBeckPRL93}. The measurement was based on an
observation that marginals of the Wigner function, which
contains complete information on the quantum state, can be
collected with the help of a balanced homodyne detector
\cite{VogeRiskPRA89}.  The Wigner function was reconstructed
from the statistics of homodyne events using a standard filtered
back-projection algorithm developed in image processing.  This
seminal work initiated an extensive research in the field of
quantum state measurement, which has brought beautiful
demonstrations of quantum phenomena as well as deeper
understanding of the foundations of quantum theory
\cite{JMOSpecialIssue}.

The Wigner function is particularly well suited for
visualization of quantum states, as it represents pictorially
quantum coherence in the form of nonclassical phase space structures.
The purpose of this Communication is to present a novel method
for reconstructing the Wigner function from homodyne statistics.
Its essential advantage compared to the standard back-projection
algorithm is the capability of compensating, in a numerically
stable way, for non-unit efficiency of the homodyne detector.
Imperfect detection is well known to have a deleterious effect
on nonclassical features of the Wigner function, such as
negativities and oscillatory interference patterns
\cite{LeonPaulPRA93,WygladzonyWigner}.  Furthermore, an attempt
to incorporate compensation into the standard linear
reconstruction scheme fails due to rapidly exploding statistical
errors and numerical instabilities
\cite{LeonPaulJMO94,DAriMaccQSO97,TanJMO97}. In the present
Communication I show that these difficulties can be effectively
overcome by taking into account {\em a priori} constraints
imposed by the quantum mechanical form of the Wigner function.
This new method is derived using the maximum-likelihood
estimation \cite{NajwiekszaWiarygodnosc}, and constraints are
formulated in a way which provides a convenient
algorithm for reconstructing the Wigner function from realistic,
finite and imperfect homodyne data.

Let us start the considerations from tracing the standard route
from raw statistics of homodyne events to the Wigner function.
The quantum mechanical probability distribution of measuring the
quadrature $x$ in a single run of the homodyne setup with the
local oscillator phase $\theta$ is given by the expectation
value of the following positive operator-valued measure
\cite{DetekcjaHomodynowa}:
\begin{equation}
\label{Eq:Hxtheta}
\hat{\cal H}(x; \theta) = \frac{1}{\sqrt{\pi(1-\eta)}}
\exp \left( - \frac{(x-\sqrt{\eta} \hat{x}_{\theta})^2}{1-\eta}
\right),
\end{equation}
where $\eta$ is the efficiency of the homodyne detector, and
$\hat{x}_{\theta}$ is the quadrature operator. In the limit
of perfect detection, the above expression reduces to
\begin{equation}
\hat{\cal H}(x; \theta) 
\; \mathrel{\mathop{\longrightarrow}_{\eta \rightarrow 1}} \;
\delta(x-\hat{x}_{\theta}),
\end{equation}
i.e., the measured statistics become distributions of quadrature
operators, which are one-dimensional projections of the Wigner
function. This projection relation can be analytically inverted,
which yields an expression for the Wigner function as the inverse
Radon transform \cite{RadonTransform} of the family of quadrature
distributions.

In standard optical homodyne tomography, regularized inverse
Radon transform is applied to frequency histograms of homodyne
events. In other words, frequency histograms are regarded as
estimates for quantum mechanical quadrature distributions, and
statistical uncertainty in the reconstruction process
is governed by a simple propagation law. This
treatment of statistical noise has crucial consequences, when
efficiency of the homodyne detector is less than one. In this
case, application of the inverse Radon transform to a family of
distributions given by Eq.~(\ref{Eq:Hxtheta}) yields a
generalized, $s$-ordered quasidistribution function with the
ordering parameter $s=-(1-\eta)/\eta$ \cite{LeonPaulPRA93}.
This object is related to the Wigner function by a convolution
with a Gaussian function. However,
the inverse relation is practically useless in numerical
calculations: linear deconvolution is an ill-posed problem,
which would enormously amplify statistical noise inevitably present in
experimental data \cite{LeonPaulJMO94}. The source of this
difficulty lies in the estimation procedure applied in standard
optical homodyne tomography. This procedure is performed at the
level of experimental frequency histograms, which are inserted in
place of quantum mechanical distributions. In such an approach,
statistical noise is regarded as necessary evil, whose impact
can be quantified, but cannot be reduced.

Fortunately, this conclusion becomes invalid if we adopt a more
careful treatment of experimental data. In fact, there is a
powerful tool for improving the performance of the
reconstruction procedure: it is {\em a priori} knowledge about
constraints satisfied by quantities to be estimated. Use of
appropriate estimation methodology allows one to incorporate
this information in the reconstruction scheme. Such an approach
can lead to substantial reduction of the statistical
uncertainty, as in this case estimates are picked only from {\em
a priori} restricted physically sensible region.  Specifically,
the quantum mechanical definition of the Wigner function results
in certain constraints on its values. The practical problem is
how to translate this {\em a priori} knowledge into an efficient
numerical algorithm. In the following, I will show that this
goal can be effectively achieved for optical homodyne
tomography.

We will start from the observation that the Wigner function at a
phase space point $(q,p)$ is given by the expectation value of
the operator $\hat{W}(q,p)$, which can be decomposed into the
diagonal form \cite{WignerandParity}:
\begin{equation}
\label{Eq:Wqp}
\hat{W}(q,p) = \frac{1}{\pi} \sum_{n=0}^{\infty} (-1)^{n}
\hat{\varrho}_{n}(q,p)
\end{equation}
where $\hat{\varrho}_{n}(q,p)$ are projections on displaced Fock states:
\begin{equation}
\hat{\varrho}_{n}(q,p) = \hat{D}(q,p) |n\rangle\langle n | 
\hat{D}^{\dagger}(q,p).
\end{equation}
Here $\hat{D}(q,p)$ denotes the displacement operator.
Quantum expectation values of $\hat{\varrho}_{n}(q,p)$ satisfy
obvious conditions:
\begin{equation}
\label{Eq:rhoconstraints}
\langle \hat{\varrho}_{n}(q,p) \rangle \ge 0, \;\;\;\;\;
\sum_{n=0}^{\infty} \langle
\hat{\varrho}_{n}(q,p)\rangle = 1
\end{equation}
The decomposition defined in Eq.~(\ref{Eq:Wqp}) suggests the
following two-step reconstruction scheme: for a given phase
space point $(q,p)$ find estimates for the family of observables
$\hat{\varrho}_{n}(q,p)$, taking into account constraints given
by Eq.~(\ref{Eq:rhoconstraints}). Using these estimates, compute
the value of the Wigner function according to
Eq.~(\ref{Eq:Wqp}). What makes this scheme more robust to
statistical noise compared to standard optical homodyne
tomography, is that estimates of $\langle \hat{\varrho}_{n}(q,p)
\rangle$ are {\em a priori} restricted to the physically
sensible region. In contrast, estimates for positive definite
observables, such as Fock state projections, obtained via
standard tomographic technique of pattern functions, may, and
often do take unphysical negative values
\cite{DAriMaccQSO97,TanJMO97}.
These artifacts become particularly strong when compensation for
detection losses is built into the pattern functions.

The positive definite estimates for $\langle \hat{\varrho}_{n}(q,p)
\rangle$ will be found using the maximum-likelihood approach.
The estimation procedure is motivated by
the one-to-one relation linking the photon distribution
with the phase-averaged homodyne statistics
\cite{MunrBoggPRA95}. This relation can be expressed
in the operator form as
\begin{equation}
\label{Eq:RandomPhase}
\frac{1}{2\pi} \int_{-\pi}^{\pi} \text{d}\theta \;
\hat{\cal H}(y;\theta) = \sum_{n=0}^{\infty} A_{n}(y) |n\rangle\langle
n|,
\end{equation}
where
\begin{equation}
A_{n}(y) = \sum_{k=0}^{n} { n \choose k}
\frac{(1-\eta)^{n-k} \eta^{k}}{\sqrt{\pi} 2^{k} k!}
H_{k}^{2}(y) \exp(-y^2)
\end{equation}
describes contribution generated by the occupation of the $n$th
Fock state. Here $H_k(y)$ is the $k$th Hermite polynomial.  The
relation given by Eq.~(\ref{Eq:RandomPhase}), along with
statistical characterization of experimental data, has been
shown to provide an algorithm for reconstructing the photon
distribution with positivity constraints \cite{BanaPRA98}.  This
approach will be now generalized to the estimation of
projections on arbitrarily displaced Fock states.

For this purpose let us apply the coherent displacement transformation to
both the sides of Eq.~(\ref{Eq:RandomPhase}), and denote the transformed
left hand side by $\hat{\cal T}(y;q,p)$. Making use of the identity
\begin{equation}
\hat{D}(q,p)\hat{x}_{\theta} \hat{D}^{\dagger}(q,p)
= \hat{x}_{\theta} - q \sin\theta - p \sin\theta
\end{equation}
yields
\begin{eqnarray}
\hat{\cal T}(y;q,p) & = &
\frac{1}{2\pi} \int_{-\pi}^{\pi} \text{d}\theta \; \hat{\cal H}
(y + \sqrt{\eta}q\cos\theta + \sqrt{\eta}p\sin\theta ;
\theta) \nonumber \\
\label{Eq:Tyqp}
& = & \sum_{n=0}^{\infty} A_n(y) \hat{\varrho}_{n}(q,p).
\end{eqnarray}
Thus, $\hat{\cal T}(y;q,p)$ as a function of $y$
describes  a probability distribution obtained by integrating the family
of {\em shifted} homodyne statistics over the local oscillator phase
$\theta$. Eq.~(\ref{Eq:Tyqp}) shows that $\hat{\cal T}(y;q,p)$
is uniquely defined by the set of projections on displaced Fock
states $\hat{\varrho}_{n}(q,p)$ and vice versa: all the quantum expectation
values $\langle \hat{\varrho}_{n}(q,p) \rangle$ can be recovered
from $\langle \hat{\cal T}(y;q,p) \rangle$. For $\eta = 1$, the 
transformation of homodyne statistics in Eq.~(\ref{Eq:Tyqp})
has a simple pictorial interpretation in
phase space: moving the phase space origin to the point $(q,p)$
corresponds to displacing the distribution of the quadrature
$\hat{x}_{\theta}$ by $q\cos\theta +p\sin\theta$.

Let us now look at the derived relation from the perspective of
raw experimental results. Data collected in $N$ runs of the homodyne
experiment consist of pairs 
$(x_i,\theta_i)$ specifying the outcome $x_i$ and the phase $\theta_i$
for an $i$th run of the setup \cite{ProblemyzFaza}.
For a given phase space point
$(q,p)$ these data can be converted to the form
\begin{equation}
\label{Eq:yi}
y_i = x_i - \sqrt{\eta} q \cos\theta_i - \sqrt{\eta} p \sin\theta_i.
\end{equation}
Statistics of these outcomes is governed by the expectation
value of $\hat{\cal T}(y;q,p)$ over the quantum state that is
measured. This probability distribution
$\langle \hat{\cal T}(y;q,p) \rangle$ contains complete
information on the set of
$\langle\hat{\varrho}_{n}(q,p)\rangle$.  However, in a real
experiment the distribution $\langle \hat{\cal T}(y;q,p)
\rangle$ is not known perfectly. The only information we have in
hand is a set of outcomes characterized by this distribution,
and {\em a priori} knowledge that 
$\langle\hat{\varrho}_{n}(q,p)\rangle$ generating it are
positive definite and sum up to one.  From this incomplete
information we need to infer estimates for projections on
displaced Fock states, which we will denote for short by
$\varrho_n$. The
solution given to this problem by the maximum-likelihood
methodology is to pick the set of $\{\varrho_n\}$ for which it
was the most likely to obtain the actual result of the series of
measurements.  Mathematically, this is done by maximization of
the log-likelihood function \cite{NajwiekszaWiarygodnosc}
\begin{equation}
{\cal L} (\{\varrho_n\};\{y_i\}) = \sum_i \ln \left(
\sum_n A_n(y_i) \varrho_n \right) - N \sum_{n} \varrho_{n},
\end{equation}
where the outcomes $y_i$ are treated as fixed parameters for 
a given phase space point $(q,p)$. In the above formula, the method of
Lagrange multipliers has been used to include the constraint 
$\sum_{n} \varrho_n = 1$.

Thus we have eventually arrived at the following computational
recipe for reconstructing the Wigner function from raw homodyne outcomes:
for a given phase space point $(q,p)$, convert experimental data
according to Eq.~(\ref{Eq:yi}) and find the maximum of the corresponding
log-likelihood function ${\cal L} (\{\varrho_n\};\{y_i\})$ over
the manifold defined by Eq.~(\ref{Eq:rhoconstraints}).
Finally, evaluate the Wigner function
as
\begin{equation}
\label{Eq:Recipe}
W(q,p) = \frac{1}{\pi} \sum_{n} (-1)^{n} \varrho_n.
\end{equation}
The nontrivial step in the above scheme is the multidimensional
constrained optimization necessary to maximize the
log-likelihood function. In solving this problem, it is
useful to note its specific form: the statistics of $\{y_i\}$
depends linearly on $\{\varrho_n\}$ which has all the properties of a
probability distribution as a function of $n$. This makes our
task a special case of linear inverse problems with positivity
constraints \cite{VardLeeJRS93}.
An effective tool in solving this class of inverse problems
is the so-called expectation-maximization (EM) algorithm
\cite{EMalgorithm}. Its
principle of operation can be understood by considering the
necessary condition for the maximum of the log-likelihood function
${\cal L}(\{\varrho_n\};\{y_i\})$. For each $m$, the partial derivative
$\partial {\cal L}/\partial \varrho_m$ must vanish, unless the maximum
is located on the boundary of the allowed region for which
$\varrho_m=0$. These two possibilities can be written jointly as
\begin{equation}
\varrho_m \frac{\partial {\cal L}}{\partial \varrho_m} = 0.
\end{equation}
This condition can be rearranged to the form
\begin{equation}
\label{Eq:FixedPoint}
\varrho_m = \frac{1}{N} \sum_i 
\frac{A_m(y_i) \varrho_m}{\sum_{n} A_n(y_i) \varrho_n}
\end{equation}
for all $m$s, which shows that the maximum likelihood estimate
for $\{\varrho_n\}$ is a fixed point of the nonlinear
transformation defined by the right hand side of
Eq.~(\ref{Eq:FixedPoint}). The idea of the EM algorithm is simply
to iterate this transformation.
When sufficient mathematical conditions are
fulfilled, this procedure converges to the maximum-likelihood
solution \cite{VardLeeJRS93}.

\begin{figure}
\epsfig{file=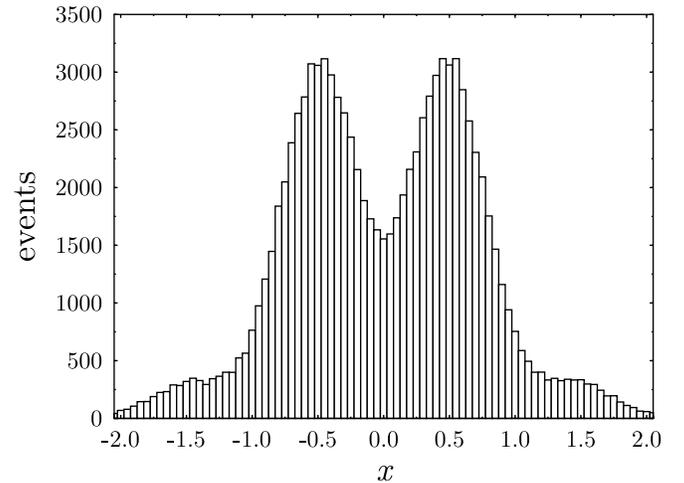,width=3.375in}
\caption{Histogram of Monte Carlo simulated homodyne events for the local
oscillator phase $\theta=0$. The modulation depth
of the interference pattern is decreased by imperfect
detection.} 
\label{Fig:Histogram}
\end{figure}

I illustrate the presented method with the reconstruction of the
Wigner function from Monte Carlo simulated imperfect homodyne
statistics for a Schr\"{o}dinger cat state
\begin{equation}
|\Psi\rangle = \frac{1}{\sqrt{2-2\exp(-2|\alpha|^2)}}
(|\alpha\rangle - |-\alpha\rangle)
\end{equation}
with $\alpha=2i$. The computer generated data consisted of
$10^5$ events simulated for each of $64$ phases uniformly spaced
between $0$ and $\pi$. The detector efficiency was assumed to
equal $\eta=90\%$. The effect of imperfect detection can be
observed in the histogram of homodyne events for the phase
$\theta=0$, depicted in Fig.~\ref{Fig:Histogram}. The visibility
of interference fringes in the homodyne statistics is
substantially smaller than in the corresponding quadrature
distribution. This blurring is a result of detection losses,
and it analogously affects the quasidistribution function
reconstructed by means of inverse Radon transform, decreasing
the magnitude of the oscillatory pattern in phase space. 

\begin{figure}
\epsfig{file=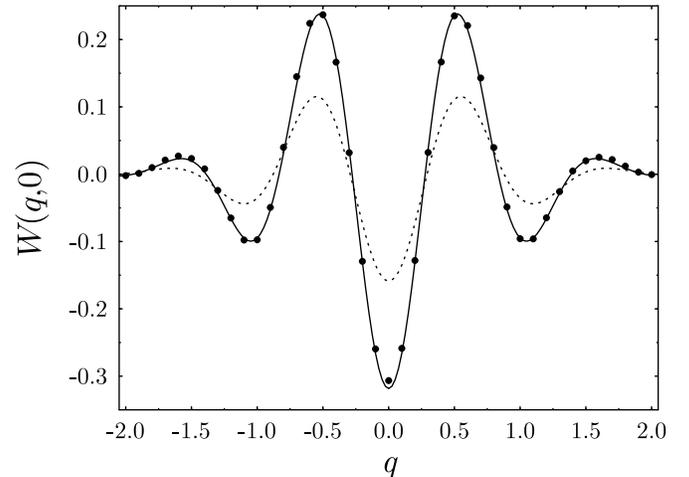,width=3.375in}
\caption{
A comparison of the reconstructed Wigner function
($\bullet$) with its analytical form (solid line). The dashed
line represents the quasidistribution function 
$\eta^{-1}W(\eta^{-1/2}q, 0;$ $ -(1-\eta)/\eta)$ that is related
via inverse Radon transform to blurred quadrature distributions
corresponding to efficiency $\eta=90\%$ \protect\cite{LeonPaulPRA93}.}
\label{Fig:Comparison}
\end{figure}

Fig.~\ref{Fig:Comparison} shows the Wigner function reconstructed 
along the position axis $q$ via $10^4$ iterations of the EM
algorithm at each point, starting from a flat distribution of $\{\varrho_n\}$
for $0\le n \le 39$. For a quantitative comparison,
the reconstructed values are plotted together
with the true Wigner function evaluated for the state
$|\Psi\rangle$. It is clearly seen that virtually full magnitude of
the oscillatory pattern is recovered despite detection losses. 
In contrast, the dashed line in Fig.~\ref{Fig:Comparison}
depicts the quasidistribution function characterized by the
ordering parameter $s=-(1-\eta)/\eta$. This function would have been 
obtained using the standard linear back-projection algorithm
from homodyne statistics in the limit of infinite number
of measurements.
 
In conclusion, the reconstruction algorithm presented in this
Communication demonstrates that application of appropriate
estimation methodology can substantially improve performance
of quantum homodyne tomography. The maximum-likelihood approach
applied in this paper provides an algorithm that is entirely
free from singularities appearing in the standard linear
reconstruction scheme. Of course, incorporation of {\em a priori}
constraints does not automatically cancel all the effects
of imperfect detection. For a fixed number of measurements, the quality
of the maximum-likelihood estimate worsens with decreasing detector
efficiency $\eta$. What is important, however, is that in any case
the obtained estimate for $\{\varrho_n\}$ remains between physical
bounds, which allows one to evaluate
safely the sum in Eq.~(\ref{Eq:Recipe}).
Quantitative discussion of statistical uncertainty in the proposed 
algorithm will be a subject of a separate publication.

It is noteworthy that the EM algorithm is a well known
tool in classical tomography and image restoration
\cite{ClassicalImaging}. However, these applications make
essential use of the fact that the multidimensional object to be
reconstructed is positive definite. This is not the case of the
Wigner function, whose negativities are a signature of
nonclassical properties. Therefore, classical reconstruction
algorithms with {\em a priori} constraints usually cannot be directly
used in quantum state measurement, and there is an apparent
need to develop novel tools for quantum tomography, such as that
presented in this paper.

{\em Acknowledgments.} The author thanks K. W\'{o}dkiewicz for
numerous comments on the manuscript, and acknowledges useful
discussions with K. Cha{\l}asi\'{n}ska-Macukow, Z. Hradil,
and J. Mostowski.
This research is supported by Komitet Bada\'{n} Naukowych,
grant 2P03B~013~15.

\end{document}